\documentclass[twocolumn,prd,superscriptaddress,
showpacs,floatfix,preprintnumbers,nofootinbib]{revtex4}

\usepackage{epsfig}
\usepackage{amsmath,bm}

\newcommand{\D}{{\rm d}}
\newcommand{\E}{{\rm e}}
\newcommand{\I}{{\rm i}}

\begin{document}

\title{Synchronization vs.\ decoherence of neutrino oscillations
at intermediate densities}

\author{Georg G.~Raffelt}
\affiliation{Max-Planck-Institut f\"ur Physik
(Werner-Heisenberg-Institut), F\"ohringer Ring 6, 80805 M\"unchen,
Germany}

\author{Irene Tamborra}
\affiliation{Max-Planck-Institut f\"ur Physik
(Werner-Heisenberg-Institut), F\"ohringer Ring 6, 80805 M\"unchen,
Germany} \affiliation{Dipartimento Interateneo di Fisica
``Michelangelo Merlin'', Via Amendola 173, 70126 Bari, Italy}
\affiliation{INFN, Sezione di Bari, Via Orabona 4, 70126 Bari,
Italy}

\date{1 December 2010}

\preprint{MPP-2010-55}

\begin{abstract}
We study collective oscillations of a two-flavor neutrino system with
arbitrary but fixed density. In the vacuum limit, modes with
different energies quickly de-phase (kinematical decoherence),
whereas in the limit of infinite density they lock to each other
(synchronization). For intermediate densities, we find different
classes of solutions. There is always a phase transition in the sense
of partial synchronization occurring only above a density threshold.
For small mixing angles, partial or complete decoherence can be
induced by a parametric resonance, introducing a new time scale to
the problem, the final outcome depending on the spectrum and mixing
angle. We derive an analytic relation that allows us to calculate the
late-time degree of coherence based on the spectrum alone.
\end{abstract}


\maketitle

\section{Introduction}                        \label{sec:introduction}

Flavor oscillations in dense neutrino gases exhibit collective
phenomena caused by neutrino-neutrino
interactions~\cite{Pantaleone:1992eq, Pantaleone:1998xi,
Samuel:1993uw, Kostelecky:1993dm, Kostelecky:1995dt, Samuel:1996ri,
Pastor:2001iu, Wong:2002fa, Duan:2005cp, Duan:2006an, Raffelt:2007yz,
Sawyer:2008zs, EstebanPretel:2007ec, Hannestad:2006nj,
Raffelt:2007cb, Raffelt:2007xt, Dasgupta:2009mg}. Recently this topic
has been studied intensely in the context of supernova neutrino
oscillations; for a recent review see Ref.~\cite{Duan:2010bg}. In
much of this literature, numerical tools were used to study cases of
practical interest for supernova neutrinos. Moreover, there has been
progress in the analytic understanding of such intriguing phenomena
as spectral splits, caused by the adiabatic decrease of the neutrino
density with radius in a supernova. However, our theoretical
understanding of many aspects of collective oscillations remains
unsatisfactory. This applies, in particular, to the question of
angular synchronization vs.\ angular decoherence of neutrinos
streaming from a supernova core~\cite{Raffelt:2007yz, Sawyer:2008zs,
EstebanPretel:2007ec}.

With such general questions in mind, we here return to an elementary
case of collective effects: synchronized oscillations in an
isotropic neutrino gas with fixed density. Modes with different
energies, $E$, and thus different vacuum oscillation frequencies,
$\omega=\Delta m^2/2E$ with $\Delta m^2$ the neutrino mass-squared
difference, quickly de-phase (kinematical decoherence). However, if
the density is large one obtains synchronized oscillations: all
modes ``stick together'' and oscillate with a common frequency
$\omega_{\rm sync} = \langle\omega\rangle$ \cite{Pastor:2001iu}. In
the language of flavor-polarization vectors ${\bf P}_\omega$, all
modes together form a common global ${\bf P}=\int\D\omega\,{\bf
P}_\omega$ that precesses as a single object.

While the extreme cases of very dilute and very dense gases are easy
to understand, the question of what exactly happens for intermediate
densities is less obvious. Different possible answers come to mind.
(i)~The system could decohere completely, but perhaps on a time scale
that grows longer for larger neutrino densities. (ii)~A stationary
state with some global ${\bf P}$ of reduced length could be reached
(partial decoherence).

The correct answer has elements of both ideas, in detail depending on
the neutrino spectrum, density, and mixing angle. In the simplest
case of maximal mixing, the system decoheres completely for densities
below a certain threshold value. For larger densities,
synchronization is partial and becomes perfect in the limit of
infinite density. The behavior of the system as a function of density
thus shows a behavior similar to a phase transition.

Other forms of behavior arise for smaller mixing angles. The main
reason is that polarization vectors that are initially aligned with
each other and almost aligned with the effective magnetic field
responsible for their precession can later deviate strongly from
both directions due to parametric resonance effects. Depending on
whether or not the resonance frequency falls within the spectrum,
complete decoherence can arise, but can take a very long time to
reach equilibrium. In other words, the decoherence time scale can be
very different from any other time scale of the problem.

Our study is in many ways complementary to Pantaleone's
investigation of the {\it Stability of incoherence in an isotropic
gas of oscillating neutrinos\/}~\cite{Pantaleone:1998xi}. He asked if
a completely incoherent initial state would spontaneously develop
global flavor polarization, a process resembling the spontaneous
synchronization of a collection of coupled oscillators. He found
neutral stability, implying that spontaneous synchronization does not
occur, largely because the equations of motions conserve energy. This
contrasts with some models designed to mimic spontaneous
synchronization of certain biological
systems~\cite{Pantaleone:1998xi}.

Returning to the polarization decay of an initially synchronized
neutrino ensemble, we begin in Sec.~\ref{sec:EoM} by setting up the
equations of motion of a system of interacting flavor polarization
vectors. In Sec.~\ref{sec:decoherence} we consider simple cases of
kinematical decoherence. In Sec.~\ref{sec:asymptotic} we derive
analytic relations that allow us to calculate the asymptotic
behavior. In Sec.~\ref{sec:maxmix} we show that the threshold
behavior of partial synchronization can be understood if one
represents the spectrum in terms of two discrete polarization
vectors. In Sec.~\ref{sec:theta} we identify the phenomenon of
self-induced resonant decoherence. In Sec.~\ref{sec:conclusions} we
summarize our findings.

\section{Equations of Motion}                          \label{sec:EoM}

\subsection{Matrices of occupation numbers}

For the kinetic evolution of a particle ensemble, the fundamental
entities are occupation numbers $f_{\bf p}=\langle a^\dagger_{\bf
p}a_{\bf p}\rangle$ for every mode~${\bf p}$. Here $a^\dagger_{\bf
p}a_{\bf p}$ is a number operator and $\langle\cdots\rangle$ an
expectation value or ensemble average. Including the flavor degree
of freedom, the occupation numbers are replaced by flavor matrices
of occupation numbers $\varrho_{\bf p,\alpha\beta}=\langle
a^\dagger_{\bf p,\alpha}a_{\bf p,\beta}\rangle$ where $\alpha$ and
$\beta$ are flavor indices. The diagonal entries represent the
occupation numbers for different flavors and are related to the survival probabilities,
whereas the off-diagonal elements encode phase information~\cite{Dolgov:1980cq,Sigl:1992fn}.
Based on this formalism, one can derive a Boltzmann collision
equation that allows one to study the evolution of a neutrino
ensemble under the influence of both flavor oscillations and
collisions~\cite{Dolgov:1980cq,Sigl:1992fn}.

Here we study a much simpler case where collisions are neglected,
motivated by neutrinos streaming freely from a supernova core.
However, in our schematic model we ignore all geometric effects and
consider the time evolution of an isotropic ensemble rather than the
spatial variation of a stationary flux. In the limit of
ultrarelativistic neutrinos, the equations of motion (EoMs)
are~\cite{Sigl:1992fn}
\begin{equation}
\I\,\dot\varrho_{\bf p}=
\left[\frac{M^2}{2E}+\sqrt{2}G_{\rm F}\,\varrho,\varrho_{\bf p}\right]\,,
\end{equation}
where $[\,{\cdot}\,,\,{\cdot}\,]$ is a commutator and $M^2$ is the
neutrino mass-squared matrix. It is not diagonal in the weak
interaction basis, causing vacuum flavor oscillations. The global
matrix $\varrho=\int\,\varrho_{\bf p}\,\D^3{\bf p}/(2\pi)^3$ has
diagonal entries of neutrino number densities. This ``neutrino
matter term'' has the same structure as the ordinary matter term
where the role of the global $\varrho$ is played by the matrix of
electron, muon and tau-lepton number densities that is diagonal in
the weak interaction basis. Indeed, a background of neutrinos would
be fully analogous to ordinary matter, were it not for the
off-diagonal elements of $\varrho$ as first recognized in Pantaleone's
seminal paper~\cite{Pantaleone:1992eq}.

Full kinematical decoherence corresponds to the off-diagonal
elements of $\varrho$ vanishing in the mass basis. At a sufficient
distance from a source, a neutrino flux will always end up in such a
state. It is sometimes described as the wave packets of different
mass eigenstates no longer overlapping.

\subsection{Polarization vectors}

For many studies it proves more transparent to represent
$\varrho_{\bf p}$ in terms of polarization vectors ${\bf P}_{\bf p}$
in ``flavor space.'' We restrict our investigation to two flavors
where $\varrho_{\bf p}$ is a Hermitean $2\times2$ matrix and thus
can be expressed in terms of Pauli matrices as $\varrho_{\bf
p}=\frac{1}{2}{\rm Tr}(\varrho_{\bf p})+\frac{1}{2}{\bf P}_{\bf
p}\cdot{\bm\sigma}$ (while for three flavors, we need Gell-Mann matrices
and 8-dimensional polarization vectors). Likewise, we may write
$M^2/2E=\frac{1}{2}{\rm Tr}(M^2/2E)+\frac{1}{2}\omega{\bf
B}\cdot{\bm\sigma}$ where $\omega=\Delta m^2/2E$ is the vacuum
oscillation frequency and ${\bf B}$ a unit vector, playing the role
of an effective magnetic field in the spin-precession formula below.

\begin{figure}[b]
\includegraphics[width=0.45\columnwidth]{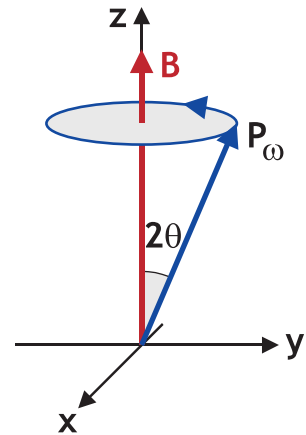}
\caption{Precession of a polarization vector ${\bf P}_\omega$ with
frequency $\omega$
around the $z$--direction (the mass direction in flavor space). The
initial orientation is in the weak-interaction direction, tilted by
twice the mixing angle relative to the
$z$--direction.\label{fig:precession}}
\end{figure}

We study an isotropic ensemble, i.e.\ we consider modes $\varrho_E$
that represent all momentum modes of the same energy $E=|{\bf p}|$.
It proves more convenient to label the modes with $\omega=\Delta
m^2/2E$ and corresponding polarization vectors ${\bf P}_\omega$. The
EoMs then appear in the familiar precession form
(Fig.~\ref{fig:precession})
\begin{equation}\label{eq:EOM1}
\dot {\bf P}_\omega={\bf H}_\omega\times {\bf P}_\omega
\quad\hbox{where}\quad
{\bf H}_\omega=\omega{\bf B}+\mu{\bf P}\,.
\end{equation}
The total polarization vector is ${\bf P} =\int\D\omega\, {\bf
P}_\omega$ which we normalize as $|{\bf P}|=1$. In this case
$\mu=\sqrt2\,G_{\rm F}n_\nu$, with $n_\nu$ the neutrino number
density, is the interaction energy felt by a neutrino in the mean
field of all others.

The precession of individual modes is around ${\bf B}$ which thus
marks a natural symmetry direction of the problem. Therefore, we
choose the ${\bf B}$--direction as the $z$--direction of a cartesian
coordinate system in flavor space. We assume that all neutrinos are
initially prepared in the same flavor, so all ${\bf P}_\omega$
initially point in the weak-interaction direction. It is tilted
relative to ${\bf B}$ by twice the mixing angle $\theta$
(Fig.~\ref{fig:precession}). Therefore, we have
\begin{equation}
{\bf B}=
\begin{pmatrix}0\\0\\1\end{pmatrix}
\quad\hbox{and}\quad
{\bf P}_\omega(0)=
\begin{pmatrix}\sin2\theta\\ 0\\ \cos2\theta\end{pmatrix}
\,g_\omega\,,
\end{equation}
where $g_\omega=g(\omega)=|{\bf P}_\omega(0)|$ is what we call the
neutrino spectrum normalized to $\int d\omega~g_\omega = 1$. The
initial direction in the $x$--$y$--plane is arbitrary, owing to a
global SU(2) symmetry of the problem: the absolute phase of $\nu_e$
relative to another flavor $\nu_x$ is arbitrary.

The EoMs conserve the quantity $\int\D\omega\,\omega{\bf B}\cdot{\bf
P}_\omega+\frac{1}{2}\,\mu\,{\bf P}^2$ which thus plays the role of
total energy. This is seen by taking the time derivative and
inserting the EoMs on the r.h.s.\ for $\dot{\bf P}_\omega$. In other
words, the EoMs can be derived from a classical Hamiltonian. This is
seen most easily if we use discrete polarization vectors ${\bf P}_i$
for a set of $N$ frequency bins $\omega_i$ where ${\bf P}=\sum_i
{\bf P}_i$,
\begin{equation}\label{eq:hamiltonian1}
H=\sum_{i=1}^N\,\omega_i\,{\bf B}\cdot{\bf P}_i+\frac{\mu}{2}\,
\sum_{i,j=1}^N{\bf P}_i\cdot{\bf P}_j\,.
\end{equation}
The polarization vectors play the role of classical angular momenta
with Poisson brackets $[P_x,P_y]=P_z$ and cyclic
permutations.\footnote{Here $[\,{\cdot}\,,\,{\cdot}\,]$ is a Poisson
bracket, not a commutator. Of course, classical variables commute in
the sense of $P_x P_y=P_y P_x$.} They inherit this property from the
SU(2) algebra of the ``flavor isospin matrices'' \cite{Duan:2006an}.
The precession equations follow from the classical Hamilton
equations of motion $\dot {\bf P}_i=[{\bf P}_i,H]$. The ${\bf P}_i$
are classical variables because they arise from occupation number
matrices which themselves are classical objects, not quantum
operators, as they represent an expectation value.

The quantity ${\bf P}$ plays the role of the total angular momentum
of the ensemble. Naturally, its component along ${\bf B}$ is
conserved. The conservation of ${\bf B}\cdot{\bf P}$ is also easily
seen by taking the time derivative of this bilinear and inserting
the EoMs for $\dot{\bf P}_\omega$.

In the polarization-vector language, our ensemble of interacting
neutrinos is equivalent to an ensemble of classical angular momenta
${\bf P}_i$ that couple to an external magnetic field with strength
$\omega_i$ and to each other with a dipole-dipole interaction of
equal strength $\mu$. However, they do not couple by some
nearest-neighbor interaction, but rather every mode couples to every
other with the same strength.

\subsection{Antineutrinos}

Occupation number matrices for antineutrinos can be defined in three
different ways, all of which have been used in the literature. If
$b^\dagger_{\bf p}$ and $b_{\bf p}$ are creation and annihilation
operators for antineutrinos in mode~${\bf p}$ we may define
$\bar\varrho_{{\bf p},\alpha\beta}=\langle b^\dagger_{\bf
p,\alpha}b_{\bf p,\beta}\rangle$ or $\langle b^\dagger_{\bf
p,\beta}b_{\bf p,\alpha}\rangle$ or $1-\langle b^\dagger_{\bf
p,\beta}b_{\bf p,\alpha}\rangle$, the latter arising from normal
ordering of $\langle b_{\bf p,\alpha} b^\dagger_{\bf
p,\beta}\rangle$. The third definition amounts to the flavor-isospin
interpretation of polarization vectors~\cite{Duan:2006an} and is the
most useful in our context. Antiparticles are interpreted as
negative-energy states and are thus described by
\hbox{$\omega=\Delta m^2/2E<0$}. In the polarization-vector
language, antineutrinos are then described by ${\bf P}_\omega$ with
negative $\omega$.

The flavor interpretation is now as follows: ${\bf P}_\omega$ in the
positive weak-interaction direction means $\nu_e$, in the negative
weak-interaction direction means $\nu_x$ if $\omega>0$, whereas for
$\omega<0$ these cases mean $\bar\nu_x$ and $\bar\nu_e$,
respectively. This sounds complicated, but is simple in the hole
interpretation of antiparticles where the presence of $\bar\nu_e$ in
mode ${\bf p}$ is equivalent to the absence of a negative-energy
$\nu_e$ from a filled mode in the Dirac sea.

Once we are studying the dynamics of our abstract Hamiltonian
problem of interacting spins, the detailed flavor interpretation is
irrelevant. We simply consider an ensemble of ${\bf P}_\omega$ with
a range of positive and negative $\omega$.

\subsection{Rotating frames}

The dynamics of the system is the same if we go to a frame rotating
with $\omega_0$ around the $z$--direction~\cite{Pantaleone:1998xi,
Duan:2005cp}. This is seen most easily if we consider the spin-spin
Hamiltonian of Eq.~(\ref{eq:hamiltonian1}) and recall that ${\bf
B}\cdot{\bf P}$ is conserved. Therefore, we may add a term
$-\omega_0{\bf B}\cdot{\bf P}$, leading to
\begin{equation}\label{eq:hamiltonian2}
H=\sum_{i=1}^N\,(\omega_i-\omega_0)\,{\bf B}\cdot{\bf P}_i+\frac{\mu}{2}\,
\sum_{i,j=1}^N{\bf P}_i\cdot{\bf P}_j\,.
\end{equation}
Therefore, even if we are having in mind an ensemble of neutrinos
(no antineutrinos) and therefore a range of positive $\omega$, we
may shift the spectrum $g(\omega)$ by a convenient amount
$-\omega_0$ and thus include negative $\omega$. In other words, the
negative-$\omega$ part of the spectrum can be interpreted as
antineutrino modes or as neutrino modes after such a shift. The
possibility of this general interpretation and the seamless
connection of neutrino and antineutrino modes is the main advantage
of the flavor-isospin interpretation of polarization vectors.

\subsection{Spectra}

We always consider non-negative spectra, $g_\omega\ge0$, avoiding
``spectral crossings.'' The latter lead to unstable forms of motion
(``flavor pendulum'' \cite{Hannestad:2006nj}) that are at the origin
of spectral splits~\cite{Dasgupta:2009mg}. Addressing the question
of decoherence in such more general cases is beyond our present
ambition.

\begin{figure}
\includegraphics[width=0.8\columnwidth]{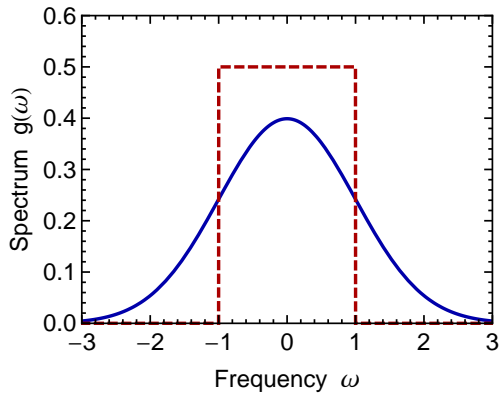}
\caption{Box spectrum (red, dashed) and Gaussian spectrum
(blue, solid) as defined in
Eqs.~(\ref{eq:Boxspectrum}) and~(\ref{eq:Gaussianspectrum}) respectively.\label{fig:spectra}}
\end{figure}

We will explicitly study two cases of simple spectra\footnote{To
simplify notation we take frequencies and time to be dimensionless,
or rather, to be expressed in terms of some fiducial frequency
$\omega_0$ so that $\omega\to\omega/\omega_0$, $\mu\to
\mu/\omega_0$, and $t\to\omega_0 t$.} that are symmetric about
$\omega=0$ and normalized to unity. One is a box spectrum where
$g_\omega=0$ everywhere except
\begin{equation}\label{eq:Boxspectrum}
g_\omega=\frac{1}{2}
\quad\hbox{for}\quad -1<\omega<+1\,.
\end{equation}
The other example is a Gaussian with unit variance
\begin{equation}\label{eq:Gaussianspectrum}
g_\omega=\frac{1}{\sqrt{2\pi}}\,
\exp\left(-\frac{\omega^2}{2}\right)\,.
\end{equation}
We show these spectra in Fig.~\ref{fig:spectra}.

The box spectrum is representative for ``compact'' spectra that are
non-zero in a finite $\omega$--range, whereas the Gaussian is
representative for spectra with long tails. We will see that
spectral tails are crucial for the phenomenon of resonant
decoherence to be discussed in Sec.~\ref{sec:theta}.

\section{Kinematical Decoherence}              \label{sec:decoherence}

\subsection{Order Parameter}

When neutrino modes de-phase relative to each other, the system
becomes in some sense disordered, but it is not trivial to define an
objective measure for this effect. Flavor oscillations alone (no
collisions) do not increase the entropy of the
system~\cite{Sigl:1992fn}. Kinematical decoherence is reversible.
For example, a neutrino flux decoheres kinematically after a short
distance from the source, but a detector with sufficient energy
resolution can still pick out oscillations in an energy-dependent
way and in this sense restores coherence. Another example is the
phenomenon of spin echo: reversing the direction of the magnetic
field unwinds the phases accrued by different modes and the original
polarization re-appears.

In our case the simplest quantity to measure coherence is what
Pantaleone calls the ``order parameter''~\cite{Pantaleone:1998xi}
\begin{equation}\label{eq:orderparameter}
R(t)=\frac{|{\bf P}_\perp(t)|}{\sin2\theta}\,,
\end{equation}
where it was assumed that $|{\bf P}(0)|=1$. Here ${\bf P}_\perp$ is
that part of ${\bf P}$ transverse to ${\bf B}$. Since ${\bf B}\cdot
{\bf P}$ is conserved, it is only the transverse component that can
shrink by kinematical decoherence. Dividing by $\sin2\theta$
normalizes the initial order parameter to $R(0)=1$.

Were we to study decoherence for more complicated spectra, this
definition would be less useful. When spectral crossings and
pendular motions are involved, ${\bf P}_\perp$ can grow even though
the system becomes ``more disordered'' as it evolves. Probably one
would need to define kinematical decoherence in a differential
sense, the dephasing of neighboring modes, leading to a shrinking of
${\bf P}_\omega$ after coarse-graining over small but finite
frequency bins $\Delta\omega$. For our simple cases we are using a
single frequency bin: the entire spectrum.

\subsection{Decoherence in the non-interacting case}

De-phasing is easy to understand in the dilute limit where
neutrino-neutrino interactions play no role ($\mu=0$) and the
evolution is determined by vacuum oscillations alone. The transverse
components evolve as $P_x(t)=\sin2\theta
\int\D\omega\,g_\omega\,\cos(\omega t)$ and
$P_y(t)=\sin2\theta\int\D\omega\,g_\omega\,\sin(\omega t)$. In other
words, the order parameter evolves as
\begin{equation}\label{eq:Fourier}
R(t)=
\left|\int\D\omega\,g_\omega\,\E^{\I\omega t}\right|\,.
\end{equation}
Therefore, $R(t)$ is given by the Fourier transform of $g_\omega$.
Unless $g_\omega$ includes ``spikes'' of the form
$\delta(\omega-\omega_i)$ at one or more frequencies $\omega_i$, the
late-time evolution is $R(t)\to0$ for $t\to\infty$ (complete
kinematical decoherence).

\begin{figure}
\includegraphics[width=0.8\columnwidth]{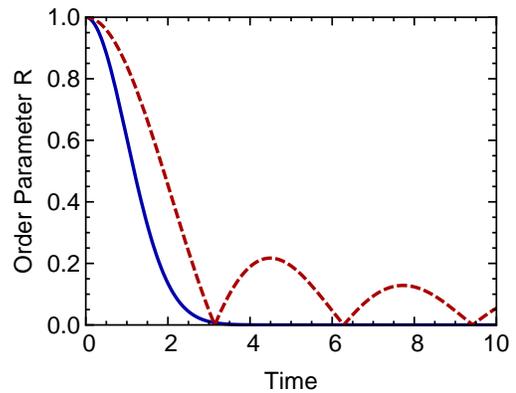}
\caption{Decay of the order parameter $R(t)$, defined in
Eq.~(\ref{eq:orderparameter}), for the box spectrum (red, dashed)
and the Gaussian spectrum (blue, solid) as given
in Eq.~(\ref{eq:boxgaussdecoherence}).\label{fig:firstexample}}
\end{figure}

For our two explicit examples, the time evolution of the order
parameter is (Fig.~\ref{fig:firstexample})
\begin{eqnarray}\label{eq:boxgaussdecoherence}
\hbox to 4em{Box:\hfil}R(t)&=&\sin(t)/t \,,
\nonumber\\
\hbox to 4em{Gauss:\hfil}R(t)&=&\exp(-t^2/2)\,.
\end{eqnarray}
The box spectrum has ``hard edges'' and thus shows the usual
``ringing'' of its Fourier transform.

\subsection{Self-maintained coherence}

\begin{figure}
\includegraphics[width=0.80\columnwidth]{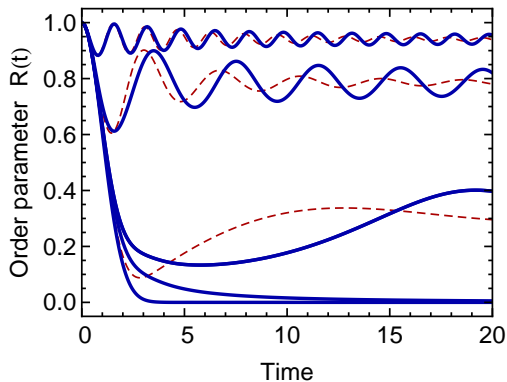}
\caption{Order parameter for the Gaussian spectrum with
maximal mixing and $\mu=0$, $\mu_0$, 1, 2 and 4 (bottom
to top), where $\mu_0\approx0.7979$ is the threshold value
for partial coherence.
Dashed lines: Solution when ${\bf P}(t)$ in the EoMs is replaced
with the asymptotic solution ${\bf A}(t)$ such as if ${\bf P}(t)$
would take on its asymptotic solution instantaneously.\label{fig:Pevol}}
\end{figure}

For non-zero interaction strength, the system can show
``self-maintained coherence'' in that an oscillatory motion persists
at late times. We illustrate this in Fig.~\ref{fig:Pevol} where we
show the evolution of the order parameter $R(t)$ for the Gaussian
spectrum with maximal mixing and different values of $\mu$. For
$\mu$ above a threshold value $\mu_0$ to be identified later, $R(t)$
at first shrinks like in the non-interacting case, but then
oscillates with decreasing amplitude around a non-vanishing
asymptotic value $R_A$. In other words, the global polarization
vector ${\bf P}(t)$ shrinks in length and eventually reaches a
nonvanishing asymptotic form ${\bf A}(t)$ that keeps precessing
around the ${\bf B}$ direction with a certain asymptotic frequency
$\omega_{A}$.  Of course, in the extreme case $\mu\to\infty$ we have
perfectly synchronized oscillations and the asymptotic length is
$|{\bf A}(t)|=|{\bf P}(0)|=1$ and $\omega_A=\omega_{\rm
sync}=\langle \omega\rangle$.

An asymptotic solution ${\bf P}(t)\to{\bf A}(t)$ performing a pure
precession around ${\bf B}$ is the simplest case of collective
motion. We may call it single-mode coherence in that the asymptotic
solution is described by a single asymptotic polarization vector.

However, we may construct more complicated spectra $g(\omega)$ where
the asymptotic solution contains two or more modes. For example, if
$g(\omega)$ consists of two separate box spectra, then for vanishing
$\mu$ the spectrum completely decoheres and for infinite $\mu$ the
system acts as one synchronized mode. However, one can choose $\mu$
so large that each individual box acts as one collective mode, being
internally synchronized, yet so small that each box essentially
precesses freely around ${\bf B}$. We now have effectively two
weakly coupled modes and thus bimodal coherence. In the same way we
can construct a hair-comb spectrum consisting of $N$ boxes, leading
for intermediate $\mu$ to $N$-modal coherence.

A priori it is not obvious if a given spectrum, for a given
interaction strength $\mu$, will decohere or not, and in the latter
case, how many modes are required to describe the asymptotic
behavior. In the rest of the paper we study only cases of
single-mode coherence where the final behavior is pure precession.
This is the case for both the box and the Gaussian spectrum.

\subsection{Sudden approximation}

We can qualitatively understand the behavior in Fig.~\ref{fig:Pevol}
if we imagine that ${\bf P}(t)$ shrinks instantaneously to its
asymptotic form ${\bf A}(t)$, an assumption that we call ``sudden
approximation.'' We can use the numerical asymptotic solution and
evolve the initial state with ${\bf A}(t)$ instead of ${\bf P}(t)$
in the EoMs, in this way reproducing what we would get if the sudden
approximation were indeed correct. We show examples as thin dashed
lines in Fig.~\ref{fig:Pevol}.

In the sudden approximation, kinematical decoherence is once more a
passive de-phasing effect except that the individual modes ${\bf
P}_\omega$ no longer precess around a common ${\bf B}$, but each
precesses around its individual ${\bf H}_\omega(t)=\omega {\bf B}
+\mu\,{\bf A}(t)$. There is no feedback of ${\bf P}_\omega$ on ${\bf
H}_\omega$ and we can determine $R(t)$ by projecting the evolution
of the individual modes on the $x$-$y$-plane. For a symmetric
spectrum with maximal mixing we find
\begin{equation}\label{eq:suddendec}
R(t)=\int\D\omega\,g_\omega\,
\frac{\kappa^2+\omega^2\cos\left(\sqrt{\kappa^2+\omega^2}\,t\right)}
{\kappa^2+\omega^2}\,,
\end{equation}
where $\kappa=\mu\,A_\perp$ which for maximal mixing is
$\kappa=\mu\,A$. For $\kappa=\mu A=0$ this reduces to
Eq.~(\ref{eq:Fourier}). For small $t$ we may expand the cosine in
Eq.~(\ref{eq:suddendec}) and find
$R(t)=1-\frac{1}{2}\,\langle\omega^2\rangle t^2$. The initial
evolution depends only on the variance of the spectrum, but not
on~$\mu$.

\section{Partial synchronization}               \label{sec:asymptotic}

\subsection{Self-consistency conditions}

We next study quantitatively the conditions for the existence of a
nonvanishing asymptotic solution ${\bf P}(t)\to{\bf A}(t)$ which we
assume to be a pure precession with frequency $\omega_A$. Conditions
for the existence of a nontrivial solution for $\mu<\infty$ can be
formulated if we study the system from the perspective of a frame
that co-rotates with ${\bf A}$ in analogy to the approach taken to
study spectral splits~\cite{Raffelt:2007cb, Raffelt:2007xt}. The
assumed final steady-state EoM is given by Eq.~(\ref{eq:EOM1}) with
\begin{equation}
{\bf H}_\omega=(\omega-\omega_A)\,{\bf B}+\mu{\bf A}\,,
\end{equation}
where in the co-rotating frame ${\bf A}$ is a static vector. The
evolution of every individual mode is now a simple precession around
this fixed vector and we refer to these asymptotic modes as ${\bf
A}_\omega(t)$.

At late times, these modes must have de-phased, meaning that their
oscillation phase relative to their ${\bf H}_\omega$ carries no
correlation with each other. Therefore, the total polarization vector
is given by the time averages of ${\bf A}_\omega$ as ${\bf
A}=\int\D\omega\,\langle {\bf A}_\omega\rangle$. The time--average of
${\bf A}_\omega$ is the projection on its Hamiltonian vector
\begin{equation}
\langle{\bf A}_\omega\rangle=
\frac{{\bf A}_\omega\cdot {\bf H}_\omega}{{\bf H}_\omega^2}
\,{\bf H}_\omega\,,
\end{equation}
implying
\begin{equation}
{\bf A}=\int\D\omega\,
\frac{{\bf A}_\omega\cdot [(\omega-\omega_A){\bf B}+\mu{\bf A}]}
{[(\omega-\omega_A){\bf B}+\mu{\bf A}]^2}
\,[(\omega-\omega_A){\bf B}+\mu{\bf A}]\,.
\end{equation}
Recall that the ${\bf A}_\omega$ in this equation are time
dependent, precessing around the asymptotic ${\bf H}_\omega$.

The quantity $P_\parallel=\cos2\theta$ is conserved and therefore
identical with $A_\parallel$. It will be useful to introduce
\begin{equation}
\kappa=\mu A_\perp
\quad\hbox{and}\quad
\omega_{\rm r}=\omega_A-\mu\cos2\theta\,,
\end{equation}
where $\omega_{\rm r}$ plays the role of a resonance frequency. The
components of the self-consistency relation perpendicular and
parallel to ${\bf B}$ become
\begin{eqnarray}\label{eq:firstconsistency}
\frac{1}{\mu}&=&\int\D\omega\,
\frac{(\omega-\omega_{\rm r})\,A_{\omega\parallel}+\kappa\,A_{\omega\perp}}
{(\omega-\omega_{\rm r})^2+\kappa^2}\,,
\nonumber\\*
A_\parallel&=&\int\D\omega\,
\frac{(\omega-\omega_{\rm r})\,A_{\omega\parallel}+\kappa\,A_{\omega\perp}}
{(\omega-\omega_{\rm r})^2+\kappa^2}\,(\omega-\omega_{\rm r}) \,,
\end{eqnarray}
where $\kappa$ and $\omega_{\rm r}$ are the quantities to be
determined.

What is the meaning of the resonance? A given ${\bf P}_\omega$
precesses in the co-rotating frame with the frequency
$\omega-\omega_A+\mu\cos2\theta$. When this quantity vanishes, ${\bf
P}_\omega$ does not precess around the $z$--direction and only feels
the transverse field. This is similar to spin resonance in a
longitudinal $B$ field and a rotating transverse field. If the
transverse field is on resonance, the spin can completely flip even
if the transverse field is weak. We will see that this resonance
effect is indeed important for small mixing angles if $\omega_{\rm
r}$ falls within the range of the spectrum.

\subsection{Sudden approximation}

The final angle between ${\bf A}_\omega$ and the corresponding
asymptotic ${\bf H}_\omega$ is not known, so we need to make
additional assumptions. Taking the non-interacting case as a guide,
decoherence happens on a timescale of the same order as the
individual oscillation frequencies and thus is not adiabatic.
Therefore, we use the sudden approximation where the initial ${\bf
P}$ shrinks instantaneously to the final ${\bf A}$. We will see that
this heuristic procedure leads to an apparently very general result
even though the sudden approximation can be badly violated.

We thus evaluate the integrand in Eq.~(\ref{eq:firstconsistency}) at
time $t=0$, assuming that at this time ${\bf A}_\omega$ coincides
with the initial ${\bf P}_\omega$ so that
$A_{\omega\perp}=g_\omega\sin2\theta$ and
$A_{\omega\parallel}=g_\omega\cos2\theta$. With the notation
$s=\sin2\theta$ and $c=\cos2\theta$ we find
\begin{eqnarray}
\frac{1}{\mu}&=&\int\D\omega\,g_\omega\,
\frac{(\omega-\omega_{\rm r})\,c+\kappa\,s}
{(\omega-\omega_{\rm r})^2+\kappa^2}\,,
\nonumber\\*
c&=&\int\D\omega\,g_\omega\,
\frac{(\omega-\omega_{\rm r})\,c+\kappa\,s}
{(\omega-\omega_{\rm r})^2+\kappa^2}\,(\omega-\omega_{\rm r})\,.
\end{eqnarray}
One can easily show that in the limit $\mu\to\infty$ the equations
are solved by $A_\perp=s$ and $\omega_A=\omega_{\rm
sync}=\int\D\omega\,g_\omega\,\omega$, the usual synchronized
oscillations.

One can form linear combinations of these equations and also use
$1=\int \D\omega\,g_\omega$. One interesting form is
\begin{equation}\label{eq:tequations}
1=\int\D\omega\,g_\omega\,
\frac{(1+t^2)\,R_A}{\xi^2+t^2R_A^2}=
\int\D\omega\,g_\omega\,
\frac{(1+t^2)\,\xi}{\xi^2+t^2R_A^2}\,,
\end{equation}
where $\xi=(\omega-\omega_{\rm r})/\mu c= (\omega-\omega_A)/\mu c+1$
and $t=\tan2\theta=s/c$. We recall that $R_A=A_\perp/\sin2\theta$ is
the asymptotic value of the order parameter with $0\leq R_A\leq1$.

\subsection{Analytic properties}

Perhaps the most illuminating form of the relations~is
\begin{eqnarray}\label{eq:scequations}
\frac{s}{\mu}&=&\int\D\omega\,g_\omega\,
\frac{\kappa}
{(\omega-\omega_{\rm r})^2+\kappa^2}\,,
\nonumber\\*
\frac{c}{\mu}&=&\int\D\omega\,g_\omega\,
\frac{(\omega-\omega_{\rm r})}
{(\omega-\omega_{\rm r})^2+\kappa^2}\,.
\end{eqnarray}
We may combine them to a single equation by adding the first,
multiplied with the imaginary unit i, to the second. Using
$(\omega-\omega_{\rm r}+\I\kappa)(\omega-\omega_{\rm r}-\I\kappa)$
for the denominator it yields
\begin{equation}
\frac{\E^{\I 2\theta}}{\mu}
=\int\D\omega\,\frac{g_\omega}{\omega-\omega_{\rm r}-\I\kappa}\,,
\end{equation}
where $\omega_{\rm r}$ is real and $\kappa$ is real and positive.

Going one step further, we add an imaginary part to $g(\omega)$  and
extend it to the complex plane by\footnote{We use capital sans-serif
letters to denote complex quantities.}
\begin{equation}
{\sf G}({\sf\Omega})=\frac{1}{\I\pi}\int_{-\infty}^{+\infty}\D\omega'\,
\frac{g(\omega')}{\omega'-{\sf\Omega}}\,.
\end{equation}
This complex-valued spectrum in the complex plane fulfills the
boundary condition $g(\omega)={\rm Re}\,{\sf G}(\omega)$ and obeys
Kramers-Kronig relations. The self-consistency relations are now
expressed in the compact form
\begin{equation}\label{eq:compact}
\mu^{-1}\,\E^{\I 2\theta}=\I\,\pi{\sf G}(\omega_{\rm r}+\I \kappa)\,.
\end{equation}
This analytic relation suggests that our heuristic argument has
provided a more general result than justified by its derivation. The
predicted asymptotic solution actually agrees with all numerical
examples even when the sudden approximation is strongly violated.

For our main examples, the extended spectra are found by direct
integration
\begin{eqnarray}\label{eq:explicitG}
\hbox to 4em{Box:\hfil}{\sf G}({\sf \Omega})&=&\frac{\I}{2\pi}\,
\log\left(\frac{{\sf \Omega}+1}{{\sf \Omega}-1}\right)\,,
\\
\hbox to 4em{Gauss:\hfil}{\sf G}({\sf \Omega})&=&
\frac{1}{\sqrt{2\pi}}\,\exp\left(-\frac{{\sf \Omega}^2}{2}\right)\,
 {\rm erfc}\left(-\I\frac{{\sf \Omega}}{\sqrt{2}}\right)\,,\nonumber
\end{eqnarray}
where ${\rm erfc}(x)=1-{\rm erf}(x)$ and erf is the error function.
The box case can be inverted, providing explicitly
\begin{equation}
\omega_{\rm r}+\I\kappa=
-{\rm coth}\left(\frac{\E^{\I2\theta}}{\mu}\right)\,.
\end{equation}

\subsection{Threshold for partial synchronization}

Even without explicit evaluation of ${\sf G}({\sf \Omega})$, we
recognize that a minimal interaction strength $\mu_0$ is required to
achieve partial synchronization. We consider vanishing asymptotic
$A_\perp$ and thus the $\kappa=0$ limit. The second integral in
Eq.~(\ref{eq:scequations}) becomes
\begin{equation}
\frac{c}{\mu}=\int\D\omega\,
\frac{g_\omega}
{\omega-\omega_{\rm 0}}\,,
\end{equation}
where $\omega_0$ is the resonance frequency corresponding to
$\kappa=0$ and the integral is to be evaluated in the
principal-value sense. Thus one can determine $\omega_0$ and use it
to evaluate the first integral where for $\kappa\to0$ the second
factor becomes $\pi\,\delta(\omega-\omega_{0})$. Therefore we find
that
\begin{equation}\label{eq:threshold}
\mu_0=\frac{\sin2\theta}{\pi\,g(\omega_0)}
\end{equation}
is required to obtain partial synchronization. For $\mu<\mu_0$ one
expects complete decoherence, although the equations may have formal
solutions with negative $A_\perp$.

\begin{figure}[b]
\includegraphics[width=0.8\columnwidth]{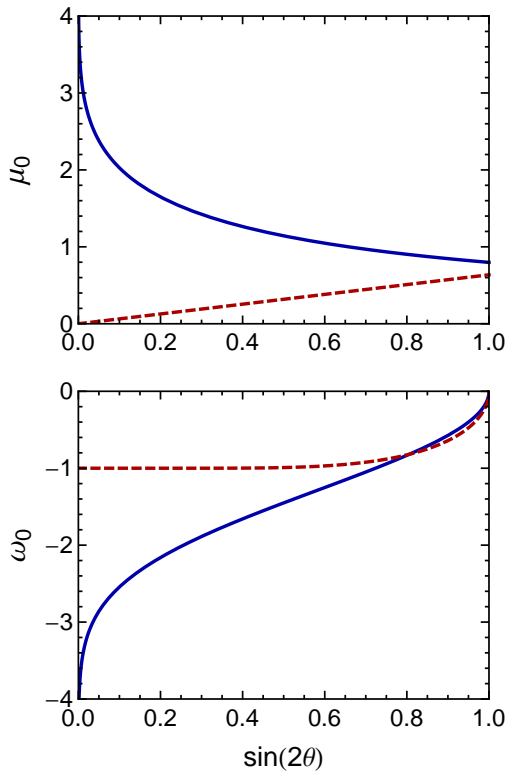}
\caption{Threshold interaction strength $\mu_0$ and threshold
resonance frequency $\omega_0$
for the Gaussian (solid blue) and the box spectrum (red dashed).
\label{fig:threshold}}
\end{figure}

For the box spectrum one finds an explicit result for any mixing
angle, whereas for the Gaussian spectrum only for maximal mixing,
\begin{eqnarray}\label{eq:explicit-thresholds}
\hbox to 4em{Box:\hfil}\mu_0&=&(2/\pi)\,s\,,
\nonumber\\
\hbox to 4em{Gauss:\hfil}\mu_0&=&\sqrt{2/\pi}
\quad\hbox{for}~s=1\,.
\end{eqnarray}
For general $s$, we use Eq.~(\ref{eq:compact}) with $\kappa=0$ and
threshold resonance frequency $\omega_0$ to obtain the parametric
solution $s_0(\omega_0)$ and $\mu_0(\omega_0)$ with
\begin{eqnarray}
s_0&=&\sin\left\lbrace{\rm arg}\bigl[\I\;{\sf G}(\omega_0)\bigr]\right\rbrace\,,
\nonumber\\
\mu_0&=&|\pi\,{\sf G}(\omega_0)|^{-1}\,.
\end{eqnarray}
Using the explicit ${\sf G}(\omega)$ for the Gaussian it is
straightforward to plot $\mu_0$ as a function of $\sin2\theta$ in
Fig.~\ref{fig:threshold}. For small $\sin2\theta$, the Gaussian case
behaves completely differently due to resonant decoherence as we
will explain in Sec.~\ref{sec:theta}.

These results for $0\leq\sin2\theta\leq1$ ($0\leq\theta\leq\pi/4$)
cover only the inverted hierarchy. In this sense it would have been
better to consider the range $0\leq\theta\leq\pi/2$ and use
$-1\leq\cos2\theta\leq+1$ as a mixing variable. For a symmetric
spectrum, however, the results are the same except that $\omega_r\to
-\omega_r$ in the normal hierarchy.

\section{Origin of Threshold}                            \label{sec:maxmix}

\subsection{Maximal Mixing and symmetric spectrum}

The component of ${\bf P}$ parallel to ${\bf B}$ is conserved and
only the transverse component decreases by decoherence. Therefore,
to understand the origin of the interaction threshold for partial
synchronization we study maximal mixing where
$P_\parallel=\cos2\theta=0$ and $\sin2\theta=1$. To simplify matters
further, we assume a symmetric spectrum $ g(-\omega)=g(+\omega)$. As
a consequence, $\omega_{\rm A}=\omega_{\rm r}=0$ for any $\mu$,
i.e.\ the global polarization vector remains static except for its
length $P(t)$ and the problem reduces to finding this function.

With $\omega_A=0$ and $\kappa=\mu A_\perp$ (which here is $\kappa=\mu
A$), we need only one real-valued self-consistency relation
\begin{equation}\label{eq:suddenintegral1}
\frac{1}{\mu}=\int\D\omega\,g_\omega\,
\frac{\kappa}{\omega^2+\kappa^2}\,.
\end{equation}
In the limit $\mu\to\infty$ we expand the second factor in
Eq.~(\ref{eq:suddenintegral1}) in powers of $\omega^2/\mu^2$ and
find $R_A=1-\langle \omega^2\rangle/\mu^2$.

\begin{figure}
\includegraphics[width=0.8\columnwidth]{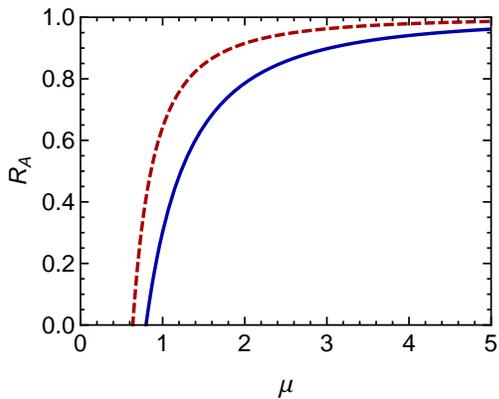}
\caption{Asymptotic order parameter
$R_A$ for the Gaussian spectrum (blue, solid)
and the box spectrum (red, dashed).\label{fig:amu}}
\end{figure}

For the Gaussian spectrum the threshold is at
$\mu_0=\sqrt{2/\pi}\approx0.7979$. The spectrum has unit variance,
so at large $\mu$ we have $R_A=1-\mu^{-2}$. Integrating
Eq.~(\ref{eq:suddenintegral1}) using Eqs.~(\ref{eq:compact})
and~(\ref{eq:explicitG}) with $\theta=\pi/4$ and
${\sf\Omega}=\I\kappa$ we find
\begin{equation}\label{eq:suddenintegral2}
\frac{1}{\mu}=\sqrt{\frac{\pi}{2}}\,
\exp\left(\frac{\kappa^2}{2}\right)
{\rm erfc}\left(\frac{\kappa}{\sqrt{2}}\right)\,.
\end{equation}
We thus have a parametric solution for $R_A(\mu)$ in the form
$\mu(\kappa)$ and $R_A(\kappa)=\kappa/\mu(\kappa)$, shown in
Fig.~\ref{fig:amu}. For the box spectrum, the threshold is at
$\mu_0=2/\pi\approx0.6366$ and
\begin{equation}
R_A=\mu^{-1}\,{\rm cotan}(\mu^{-1})\,.
\end{equation}
We show this result as a red, dashed line in Fig.~\ref{fig:amu}.

\subsection{Discrete polarization vectors}        \label{sec:bimodal}

When the system develops a collective mode, it settles only slowly
to the final state of a pure precession, maintaining a long-lasting
collective oscillation around the asymptotic solution. This
observation suggests that the phase transition to a non-vanishing
average $\langle P\rangle$ may already occur in a simpler system. To
investigate this question we represent the spectrum schematically by
two discrete polarization vectors, one for the negative and one for
the positive frequency modes, respectively. To be specific, we use
the normalized symmetric spectrum
\begin{equation}
g_\omega=\frac{\delta(\omega+\frac{1}{2})+\delta(\omega-\frac{1}{2})}{2}\ ,
\end{equation}
and maximal mixing. This coarse representation of the box spectrum
does not allow for decoherence, yet we may study its evolution and
ask for its eigenmodes as a function of the interaction strength.

\begin{figure}
\includegraphics[width=0.8\columnwidth]{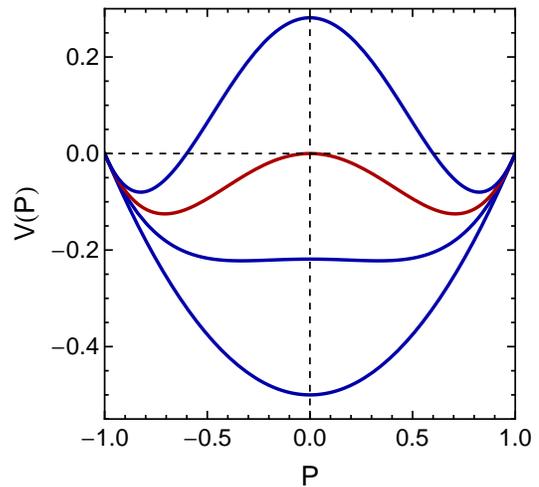}
\caption{Potential $V(P)$ defined in Eq.~(\ref{eq:potential})
for $\mu=0$, $\frac{3}{4}$, 1, and $\frac{5}{4}$ (bottom to top).
\label{fig:potential}}
\end{figure}

The symmetries of our problem imply that ${\bf P}(t)$ has only one
nonvanishing component $P(t)$. From the full EoMs we find an
equation of the form ``kinetic plus potential energy is constant''
$\frac{1}{2}\,{\dot P}^2+V(P)=0$ with
\begin{equation}\label{eq:potential}
V(P)=\frac{\mu^2(1-P^2)^2-(1-P^2)}{2}\,.
\end{equation}
We show this potential in Fig.~\ref{fig:potential} for several
values of $\mu$.

\begin{figure}
\includegraphics[width=0.8\columnwidth]{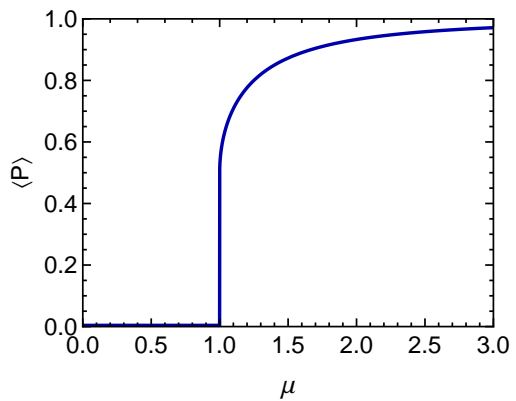}
\caption{$\langle P\rangle$ given by Eq.~(\ref{eq:as2}) for the case of
two discrete polarization vectors.\label{fig:zav}}
\end{figure}

The motion starts at $P=1$ with total energy 0. For $0\leq\mu<1$ the
motion oscillates between $-1\leq P\leq+1$ and $\langle P\rangle=0$.
On the other hand, for $\mu>1$ the system is trapped in the region
of positive $P$. It oscillates in the range $P_{\rm min}<P<1$ around
the average \hbox{$\langle P\rangle=\frac{1}{2}(1+P_{\rm min})$}.
The turning point is $P_{\rm min}=\sqrt{1-\mu^{-2}}$ so that
\begin{equation}\label{eq:as2}
\langle P\rangle=\frac{1+\sqrt{1-\mu^{-2}}}{2}\,.
\end{equation}
For $\mu<0$ we have $\langle P\rangle=0$ and thus we find the
overall behavior shown in Fig.~\ref{fig:zav}.

The simple system of two discrete polarization vectors nicely
reproduces the key features observed for a continuous spectrum. In
particular, the threshold behavior is primarily related to the
global properties of the self-interacting system. Decoherence toward
the asymptotic $P(t)$ is a secondary effect.

\section{Resonant Decoherence}                      \label{sec:theta}

\subsection{Vanishing mixing angle}

\begin{figure}
\includegraphics[width=0.8\columnwidth]{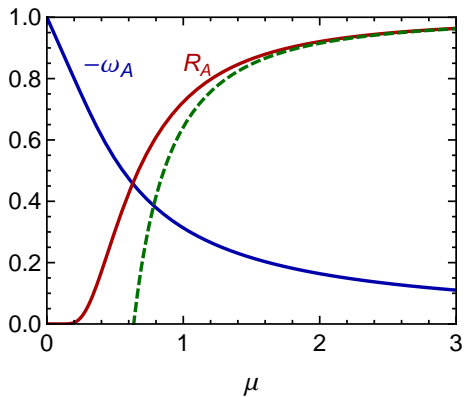}
\caption{Asymptotic values for global precession frequency $\omega_A$
(blue)
and order parameter $R_A$ (red), assuming our
box spectrum with vanishing mixing angle. The green (dashed) curve is
$R_A$ for maximal mixing.\label{fig:wa}}
\end{figure}

For non-maximal mixing, new phenomena can arise. We continue to use
a symmetric spectrum $g_\omega$, but the co-rotation frequency for
$\mu<\infty$ is no longer identical with the synchronization
frequency $\omega_{\rm sync}=0$. As an extreme case we consider the
limit of vanishing mixing angle. One may begin with the
self-consistency conditions in the form of
Eq.~(\ref{eq:tequations}). In the limit $\theta\to 0$, up to linear
order in $\theta$, we have $t^2=\tan^2(2\theta)\to0$ and
$c=\cos2\theta\to1$ and the conditions become
\begin{equation}\label{eq:sumrules-zerotheta}
1=\int\D\omega\,g_\omega
\frac{\mu^2\,R_A}{(\omega-\omega_{\rm r})^2}
\quad\hbox{and}\quad 1=\int\D\omega\,g_\omega\,
\frac{\mu}{\omega-\omega_{\rm r}}\,.
\end{equation}
The second integral does not depend on $R_A$ and thus can be solved
and inverted to obtain $\omega_{\rm r}(\mu)$. The result can be
inserted in the first integral to find $R_A(\mu)$. For the box
spectrum we find explicitly
\begin{equation}\label{eq:wa-zerotheta}
\omega_{\rm r}(\mu)=-{\rm coth}(\mu^{-1})
\quad\hbox{and}\quad
R_A(\mu)=\frac{{\rm coth}^2(\mu^{-1})-1}{\mu^2}\,.
\end{equation}
These results also follow from Eq.~(\ref{eq:explicitG}) in the $s\to
0$ limit. We show $-\omega_{\rm A}(\mu)$ and $R_A(\mu)$ in
Fig.~\ref{fig:wa}. Note that $-1<\omega_{\rm A}<0$ and in particular
$\omega_A\to \omega_{\rm sync}=0$ for $\mu\to\infty$. For small
$\mu$ the function $R_A(\mu)$ is very small, but there is no
threshold.

\subsection{Self-induced decoherence}

Inspecting Eq.~(\ref{eq:sumrules-zerotheta}) we notice that the
first integral diverges if $g_\omega>0$ for $\omega=\omega_{\rm r}$,
whereas the second remains meaningful in the principal-value sense.
In other words, $\omega_{\rm r}$ can be determined from the second
equation, whereas the first implies $R_A=0$, i.e.\ complete
decoherence for any $\mu$. This conclusion applies to any spectrum
with infinite tails, whereas the resonance frequency is outside of
the box spectrum for any $\mu$.

To understand better the impact of the resonance, we return to the
symmetric normalized box spectrum, augmenting it with small
``wings''
\begin{equation}\label{eq:boxwings}
g_\omega=\begin{cases}(1-b)/2&\hbox{for}~|\omega|<1\\
b/2&\hbox{for}~1<|\omega|<2\\
0&\hbox{otherwise}\end{cases}
\end{equation}
where $0<b\ll 1$. The self-consistency conditions imply
qualitatively different solutions for any nonzero $b$ if
$\omega_{\rm r}$, which is essentially fixed by the main part of the
box, falls within the wings. On the other hand, if $b\ll1$ the
evolution of $P(t)$ can not at first differ from the case $b=0$, so
the impact of a small $b$ can only manifest itself at late times.

\begin{figure}
\includegraphics[width=0.8\columnwidth]{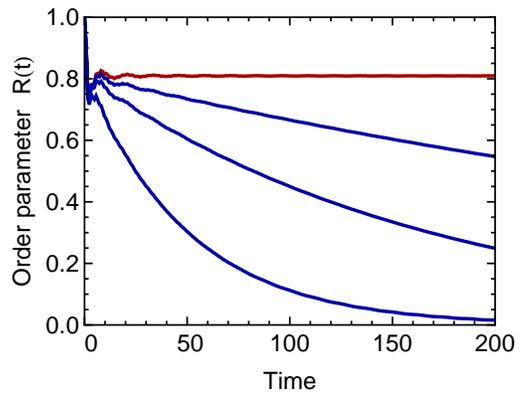}
\caption{Evolution of the order parameter for
the box spectrum with wings of Eq.~(\ref{eq:boxwings}), using $b=0$,
$10^{-3}$, $3\times10^{-3}$, and $10^{-2}$ (top to
bottom).\label{fig:Pboxres}}
\end{figure}

We study numerically an example where the resonance frequency is
fixed to fall into the center of the negative-frequency wing, i.e.\
$\omega_{\rm r}=-1.5$, implying with Eq.~(\ref{eq:wa-zerotheta})
$\mu=1/{\rm acoth}(3/2)=2/\log(5)\approx1.243$ for $b\ll1$. The
evolution of $R(t)$ is shown in Fig.~\ref{fig:Pboxres}. For
nonvanishing $b$ it is exponentially damped, the rate depending
on~$b$.

\begin{figure}[b]
\includegraphics[width=0.8\columnwidth]{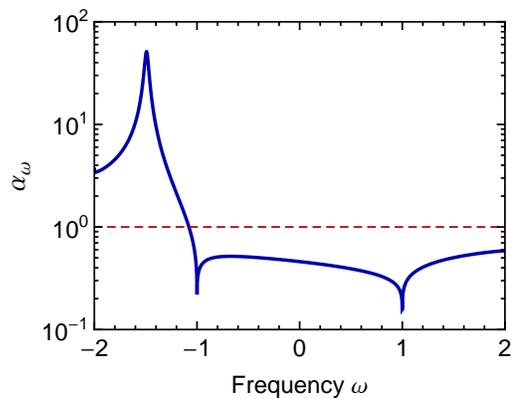}
\caption{Final deviation of polarization vectors from the ${\bf B}$
direction in units of $\sin2\theta$. The box spectrum
with wings was used with $b=10^{-2}$. In the sudden approximation one
would expect $\alpha_\omega=1$ (dashed line).\label{fig:Pboxangles}}
\end{figure}

We also consider the late-time orientation of ${\bf P}_\omega$
relative to ${\bf B}$. When ${\bf P}_\perp$ has shrunk to zero, the
remaining motion of ${\bf P}_\omega$ is a precession around the
$z$--direction with a fixed angle. In Fig.~\ref{fig:Pboxangles} we
show $\alpha_\omega=P_{\omega,\perp}/(g_\omega\,\sin2\theta)$, a
quantity that is initially unity when all polarization vectors begin
in the weak-interaction direction. We see that indeed the
polarization vectors near $\omega_{\rm r}=-1.5$ have moved far away,
whereas those with larger frequencies have actually been pulled
closer to ${\bf B}$. We show the result for $b=10^{-2}$. For other
values the curve looks the same, except that the resonant peak is
taller for smaller $b$. For other spectra (e.g.~Gaussian), the curve
is different in detail, but there is always the resonance peak.

\subsection{Intermediate mixing angle}

When the mixing angle is not vanishingly small, the polarization
vectors cannot be driven arbitrarily far away from the ${\bf B}$
direction. For example, ${\bf P}_\omega$ exactly on resonance will
completely reverse and then come back and so forth. For example, the
resonance peak in Fig.~\ref{fig:Pboxangles} reaches roughly up to
$P_{\omega,\perp}=50\,g_\omega\,\sin2\theta$. Therefore, if
$\sin2\theta>0.02$, it is no longer geometrically possible to get
that far and the resonance can not fully develop.

The threshold condition for partial synchronization of
Eq.~(\ref{eq:threshold}) here reads $s_0=\pi\mu b/2$, assuming $\mu$
is such that the resonance falls into the wing. Specifically we
consider $\mu=2/\log(5)\approx1.243$ and $b=10^{-2}$. With these
numbers we expect complete decoherence for $\sin2\theta<0.01952$, in
perfect agreement with our simple geometric estimate.

\begin{figure}[b]
\includegraphics[width=0.8\columnwidth]{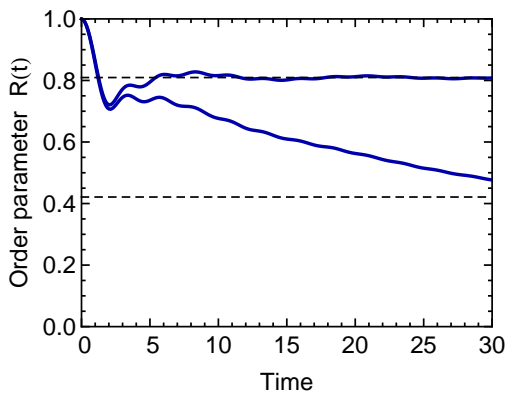}
\caption{Evolution of order parameter for box spectrum with
$\sin2\theta=0.04$. Top curve: Simple box ($b=0$). Bottom curve:
$b=0.01$. Dashed lines show the asymptotic values.\label{fig:amedmix}}
\end{figure}

The different forms of evolution are illustrated in
Fig.~\ref{fig:amedmix}, showing the early evolution for the box with
wings ($b=0.01$ and $\sin2\theta=0.04$) in comparison with the
simple box. The evolution is a two-step process, quickly decohering
kinematically to what would be the final level for the simple box,
and then decaying exponentially to the final level relevant when the
spectrum has a wing.

\begin{figure}
\includegraphics[width=0.8\columnwidth]{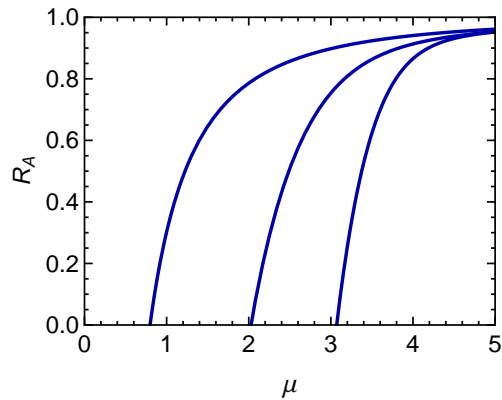}
\caption{Asymptotic order parameter for Gauss spectrum with mixing angles
$\sin2\theta=1$, 0.1 and 0.01 (left to right).\label{fig:gaussthreshold}}
\end{figure}

For a spectrum with a long tail such as the Gaussian spectrum, the
resonance always falls into the spectral range. In the
$\sin2\theta=0$ limit one can not obtain stable synchronized
oscillations---eventually the system decoheres, no matter how large
$\mu$, although the exponential decay is very slow for large $\mu$.
For nonvanishing mixing there exists a threshold so that for
$\mu>\mu_0$ partial synchronization arises. In
Fig.~\ref{fig:gaussthreshold} we show the Gaussian $R_A(\mu)$ for
several mixing angles. In contrast to the box spectrum, there is
always a threshold and it moves to larger $\mu$ as expected.

\section{Conclusions}                          \label{sec:conclusions}

We have investigated the most elementary case of collective neutrino
oscillations, an isotropic gas prepared in a single flavor with a
fixed density. The seemingly simple question of partial
synchronization for a large but finite density reveals a rich
variety of possible solutions.

The idea of a late-time asymptotic solution has led us to formulate
self-consistency relations that can be evaluated in the sudden
approximation and that provide the asymptotic solution from the
spectrum alone. We find excellent agreement of the asymptotic
solutions such predicted with those from a numerical solution of the
full EoMs. The agreement persists even when the true solution does
not resemble the sudden approximation and we suspect that these
relations are exact.

Partial synchronization requires a minimum density (a minimum
effective interaction strength $\mu$). For smaller densities,
decoherence is complete. For larger $\mu$, synchronization is
partial and becomes perfect for $\mu\to\infty$. This ``phase
transition behavior'' derives from a simple model where one
coarse-grains the spectrum in terms of two polarization vectors,
representing two halves of the spectrum. The system is represented
by a particle moving in a double-well potential, and depending on
$\mu$ it is trapped in one well or traverses both.

Non-maximal mixing enables a parametric resonance where those
polarization vectors near the resonance frequency are driven away
from their initial orientation. The back-reaction on the system
consists of decoherence. This effect can only happen if the spectrum
actually includes the resonance frequency, but this is always the
case for spectra with long tails. The exact outcome depends on the
spectrum and the mixing angle.

Decoherence can thus arise in different ways. In the dilute limit it
consists of purely kinematical de-phasing of different modes. In the
interacting system the effect is similar, except that each mode
precesses around a different direction and this direction changes as
the system decoheres. The self-induced resonant decoherence effect
that we have discovered, on the other hand, is dynamical in the
sense that an internal resonance of the system leads to an
exponential decay of the global polarization vector. This decay can
be slow, depending on parameters, and introduces a new time scale.

Our study is complementary to Pantaleone~\cite{Pantaleone:1998xi}
who used the completely incoherent state as an initial condition.
Based on an analytic argument, Pantaleone found that such a system
will never develop spontaneous polarization (``neutral stability'').
He also argued that a collective mode can be supported for any
strength of $\mu$, in seeming contrast to our finding of a minimum
required $\mu$ to maintain partial coherence. However, this question
depends on the assumed initial condition. Beginning with all
polarization vectors aligned with each other (our case), indeed a
minimum $\mu$ is required for a coherent motion to survive. On the
other hand, one can set up a pure precession mode for any strength
of $\mu$ where all polarization vectors are arranged in a
co-rotating plane, not along a common direction
\cite{Raffelt:2007cb, Raffelt:2007xt}. In practice this can be
achieved by beginning with $\mu\to\infty$ (synchronized limit) and
then reduce $\mu$ adiabatically to any desired value. This procedure
leads to a spectral split if $\mu$ is reduced all the way to zero,
and to a pure precession mode for any non-vanishing $\mu$.

\section*{Acknowledgements} 

This work was partly supported by the Deutsche
Forschungsgemeinschaft under grant TR-27 ``Neutrinos and Beyond,''
the Cluster of Excellence ``Origin and Structure of the Universe''
(Munich and Garching), and by the Italian MIUR and INFN through the
``Astroparticle Physics'' research project.


\end{document}